\documentclass[aps, prb, twocolumn, showpacs,amsmath,amssymb]{revtex4}
\usepackage{graphicx}
\usepackage{bm}
\begin{document}

\title{On the valence bond solid in the presence of  Dzyaloshinskii-Moriya interaction}
\author{Chen-Nan Liao}
\author{Chyh-Hong Chern}
\email{chchern@ntu.edu.tw}

\affiliation{Department of Physics, National Taiwan University, Taipei 10617, Taiwan}

\begin{abstract}
We examine the stability of the valence bond solid (VBS) phase against the Dzyaloshinskii-Moriya (DM) interaction in the bipartite lattice.  Despite the VBS is vulnerable against the antiferromagnetic interaction, for example in the Q-J model proposed by Sandvik, where the quantum phase transition occurs at $J^*/Q = 0.04$, we found that on the contrary the VBS is very stable against the DM interaction.  The quantum phase transition does not occur until D/Q goes to infinity, where D is the strength of the DM interaction.  The VBS in the ALKT model and the Haldane gap system also exhibit the same property.
\end{abstract}

\pacs{75.10.Pq, 75.30.Gw, 75.40.Mg} 
\maketitle

Valence bond solid (VBS) is the array of singlets that all the antiferromagnetic spins are paired statically with one of their nearest-neighbor spins in the real space.  It preserves the spin rotational symmetry because all spins are paired in singlets and consequently the total magnetization is zero.  Since the spins do not change their pairing partners , those singlets usually form a periodic array that breaks the translational symmetry in addition to the lattice symmetry.  Its "liquid phase", usually recognized as the resonating valence bond (RVB)\cite{Pauling1968, Anderson1987}, restores the translational symmetry by forming the dynamical singlets that spins can change their pairing partners dynamically with the infinite or the short range pairing strength.  Both of them have attracted great attention to the condensed matter physicists because of their unusual ground state and the excitation properties.  VBS usually accompanies with a spin gap by breaking one pair of singlets.  The RVB, on the other hand, often requires the critical spin correlation, and interestingly the fundamental excitations may carry the fractional quantum number. 

Albeit they are immensely interesting, pure spin models to host them as the ground state is not easy to construct, especially in the dimensions greater than one.  Therefore, their ground state as well as the excitation properties are not yet well understood.  Part of the reasons is that the antiferromagnetic energy often dominates when competing with the VBS and the RVB states.  Consequently, the ground state favors the antiferromagnetic (AFM) state rather than the VBS or the RVB states.  

Since we are interested in the properties of the VBS, the model Hamiltonians to host it as the ground state are necessary.  The easiest way to realize the VBS ground state is the one-dimensional spin Peierls system, where a soft lattice is required as well as the strong spin-lattice interaction.  The spin Peierls system does not serve our satisfaction, because we are looking for the VBS state from a pure spin model with low spin quantum number at best.  Another possible system with the VBS ground state in one dimension is the Haldane gap system, namely the Heisenberg model with integer spins.  One realization of the Haldane gap system is the AKLT model\cite{affleck1988cmp} described by the following Hamiltonian
\begin{eqnarray}
H_{\small AKLT} = J\sum_{i}[\vec{S}_i\cdot\vec{S}_{i+1}+\frac{1}{3}(\vec{S}_i\cdot\vec{S}_{i+1})^2+\frac{2}{3}], \label{eq:AKLT}
\end{eqnarray}
with $S=1$ and $J>0$.  The spin one on each site is constructed by the symmetric projection of two spin-1/2 spins.  Since the coordination number is two, these two spin-1/2 spins on every site form the singlets with each nearest neighbor.  This state is certainly the VBS state and can be proved to be the exact ground state of Eq.(\ref{eq:AKLT}).

In two dimensions, there are few spin models that exhibit VBS ground states.  For example, the SU(N) Heisenberg model is shown to have the columnar VBS state\cite{Read1989npb, Read1990prb, kawashima2007prl} for N $\ge 5$ and the AFM for N $< 5$.  Recently, Sandvik\cite{Sandvik2007prl} proposed a ring-exchange-like model and considered the quantum phase transition between the VBS and the AFM, described by the following Hamiltonian
\begin{eqnarray}
H_{QJ} = -Q\!\sum_{<ijkl>}\!(\vec{S}_i\!\cdot\!\vec{S}_{j}\!-\!\frac{1}{4})(\vec{S}_k\!\cdot\!\vec{S}_{l}\!-\!\frac{1}{4})+J\!\sum_{<ij>}\!\vec{S}_i\!\cdot\!\vec{S}_{j} \label{eq:QJ}
\end{eqnarray}
where $<ij>$ denote the nearest neighbor spins and $<ijkl>$ are the coordinates of a plaquette with the counterclockwise or the clockwise order.  We shall call it QJ model in the rest of the paper. Using this model, Sandvik found that a second order quantum phase transition from the VBS state to the AFM state take place at $J/Q = 0.04$.  This small ratio of $J/Q$ indicates that the VBS state is indeed vulnerable to the AFM perturbation.  Another model that might host the VBS ground state is the Heisenberg model in the kagome lattice\cite{Singh2007prb}.  Numerical calculation showed that it is a VBS with 36-site  unit cell.  Although the final verdict of the ground state of the Heisenberg model in the kagome lattice is still under debate, this model is worth mentioning, because it relates to the ZnCu$_3$(OH)$_6$Cl$_2$ that shows no spin ordering down to tens of the mK as well as the absence of the spin gap down to 0.1meV\cite{Helton2007prl}.  The lack of the spin ordering is also one of the important properties of the VBS state.  However, since there is a spin gap in the VBS system, the Heisenberg model can not fully describe the material.

In this paper, we study the Dzyaloshinskii-Moriya (DM) perturbation in the VBS state.  Part of the motivation was enlightened by the experimental fact that VBS systems often occur in the lattice with low crystal symmetry for example the Ni(C$_2$H$_8$N$_2$)$_2$NO$_2$(ClO$_4$) (NENP) for the Haldane gap system and the ZnCu$_3$(OH)$_6$Cl$_2$.  Especially, if the inversion symmetry is absence, the DM interaction is often inevitable.  Another motivation is tempted by the study of ZnCu$_3$(OH)$_6$Cl$_2$.  Further numerical study had considered the addition of the DM perturbation to the Heisenberg model in the kagome lattice.  Because the kagome lattice does not have the inversion center, the DM interaction must play some role in this material.  ESR measurement detect the strength of the DM interaction is roughly 0.08J\cite{Zorko2008prl}.  On the other hand, theoretical study 
\cite{Cepas2008prb} suggests that there is a quantum phase transition from the VBS state to the AFM state at $D/J = 0.1$.  If it is true, the VBS state would be also as vulnerable against the DM perturbation as well as the AFM one.  Regardless the inconsistency of their results with previous studies that strikes our understanding of ZnCu$_3$(OH)$_6$Cl$_2$, the question "does the vulnerability of the VBS by the DM interaction also happen in other VBS systems?" has its own weight, which will be addressed in this paper.  

To get the general understanding to that question, we consider the VBS states in the AKLT and the QD models, where the QD model is meant by letting J = 0 and adding the DM term in the QJ model.  These two models are chosen, because they are of the low-spin models and there is almost no doubt that the ground state in these two models is the VBS state.  Our results, on the contrary, show the stability of the VBS states against the DM interaction in those two models.  The quantum phase transition does not occur at any finite $D$.  We computed the spin gap and found that it does not close until D/Q  in the QD model (or D/J in the AKLT model) goes to infinity, where $D$ is the strength of the DM interaction.

Our results are consistent with the previous study of the XXZ Heisenberg model of spin one.   Consider the Haldane gap system with the DM interaction
\begin{eqnarray}
H = J\sum_{i}\vec{S}_i\cdot\vec{S}_{i+1} + \sum_i \vec{D}_i\cdot (\vec{S}_i\times\vec{S}_{i+1}). \label{eq:Haldane+DM}
\end{eqnarray}
If we choose $\vec{D}\parallel \hat{z}$, Eq.(\ref{eq:Haldane+DM}) becomes
\begin{eqnarray}
H = \sum_i [J_\bot S^+_iS^-_{i+1}+h.c.+JS^z_iS^z_{i+1}],
\end{eqnarray}
where $J_\bot = J+iD_i$.   It implies that the DM term can be gauged away and maps to the XXZ Heisenberg model
\begin{eqnarray}
H_{\small XXZ} = \sum_i(S^x_iS^x_{i+1}+S^y_iS^y_{i+1}+\lambda S^z_iS^z_{i+1}). \label{eq:XXZ}
\end{eqnarray} 
with $\lambda < 1$.  Obviously, the SO(3) symmetry is explicitly broken to U(1) for a finite $D$.  In this case, one might expect that the quantum phase transition to the XY-phase might occur at some $\lambda^*$.  However, numerical study\cite{Neirotti1999prb} showed that the quantum phase transition, which is shown to be of the Kosterlitz-Thouless type, does not happen until $\lambda =0$.  In other words, only infinitely strong DM interaction can close the spin gap and destroy the VBS state.  Therefore, very unlike the vulnerability of the questionable VBS state in the Heisenberg model in the kagome lattice,  the possible VBS state in the Haldane gap system is very stable against the DM interaction.  The key lies in the coincidence that the DM term can be gauged away and the spin gap is robust at any finite $\lambda \le 1$ in the XXZ Heisenberg model.

Since the AKLT model was proposed to realized the Haldane gap, we expect that the same results will happen, and indeed it is.  We also examine the VBS state in the QD model and find the same stability.  One might speculate that the vulnerability of the VBS in the Heisenberg model in the kagome lattice
might be due to the incapability of the DM interaction to be gauged away.  In the AKLT and the QD model, the DM interaction can not be gauged away either.  If the VBS state in the Heisenberg model in the kagome lattice is proven to be true by more detail examination, our results may imply that the VBS state in that case is very different from the ones we study here.  Therefore, this exercise might shed light on the characterization of the VBS states.

A simple-minded explanation of these results can be understood as the following.  Consider the case that there are only two spins with the interaction $J\vec{S}_1\cdot\vec{S}_2$ and $J>0$.  The ground state of this case is the singlet state with the energy $-3J/4$, and the excited state is the triplet with the energy $J/4$.  Introducing the DM interaction of $\vec{D}\parallel \hat{z}$ rearranges the eigenvalues to be $(J/4, J/4,1/4(-J-2\sqrt{D^2+J^2}), 1/4(-J+2\sqrt{D^2+J^2}))$.  Two unaffected states are $|\uparrow\uparrow>$ and $|\downarrow\downarrow>$ in the triplet.  It is obvious to see that introducing any finite $D$ does not cause level crossing.  In other words, the spin gap does not close at any finite $D$ value\footnote{We note that the result for $D=\infty$ disconnects the one for the finite D, since in the former case $J$ can be take to be zero}.  

We organize this paper in the following way.  In the section \ref{section:AKLT}, we applied the DMRG method to compute the spin gap as the function of D/J in the AKLT plus DM interaction.  In section \ref{section:QD}, we use the Lanczos method to compute the spin gap in the QD model mentioned above.  Finally, we conclude our results in the section \ref{section:CD}.

\section{AKLT model} \label{section:AKLT}
In this section, we compute the spin gap of the AKLT model plus the DM interaction given by
\begin{eqnarray}
H_{\small AKLT+DM} = H_{\small ALKT} + H_{\small DM},
\end{eqnarray}
\begin{eqnarray}
H_{\small DM} = \sum_i \vec{D}_i\cdot (\vec{S}_i\times\vec{S}_{i+1}). \label{eq:DM}
\end{eqnarray}
We consider the uniform DM vector $\vec{D}\parallel \hat{z}$ here, where $\vec{z}$ is perpendicular to the chain direction.  The result of the stagger DM vector will be the same.  As mentioned above, the nearest-neighbored DM interaction can be gauged away in the Heisenberg model, but it is not true in the AKLT model because of the $(\vec{S}_i\cdot\vec{S}_{i+1})^2$ term.  If we consider only the Eq.(\ref{eq:DM}) at the classical level, an magnetic ordered state with the $4a$ periodicity will be favored as the ground state, where $a$ is the lattice constant.  Similar to the QJ model, the quantum phase transition from the VBS state to the DM ground state might be expected.  To resolve this, we apply the DMRG method to compute the spin gap with the $m$ to be used around 400.  We solve the problem using the open boundary condition, and the system size goes up to 300 spins.

\begin{figure}[htb]
\includegraphics[width=0.5\textwidth]{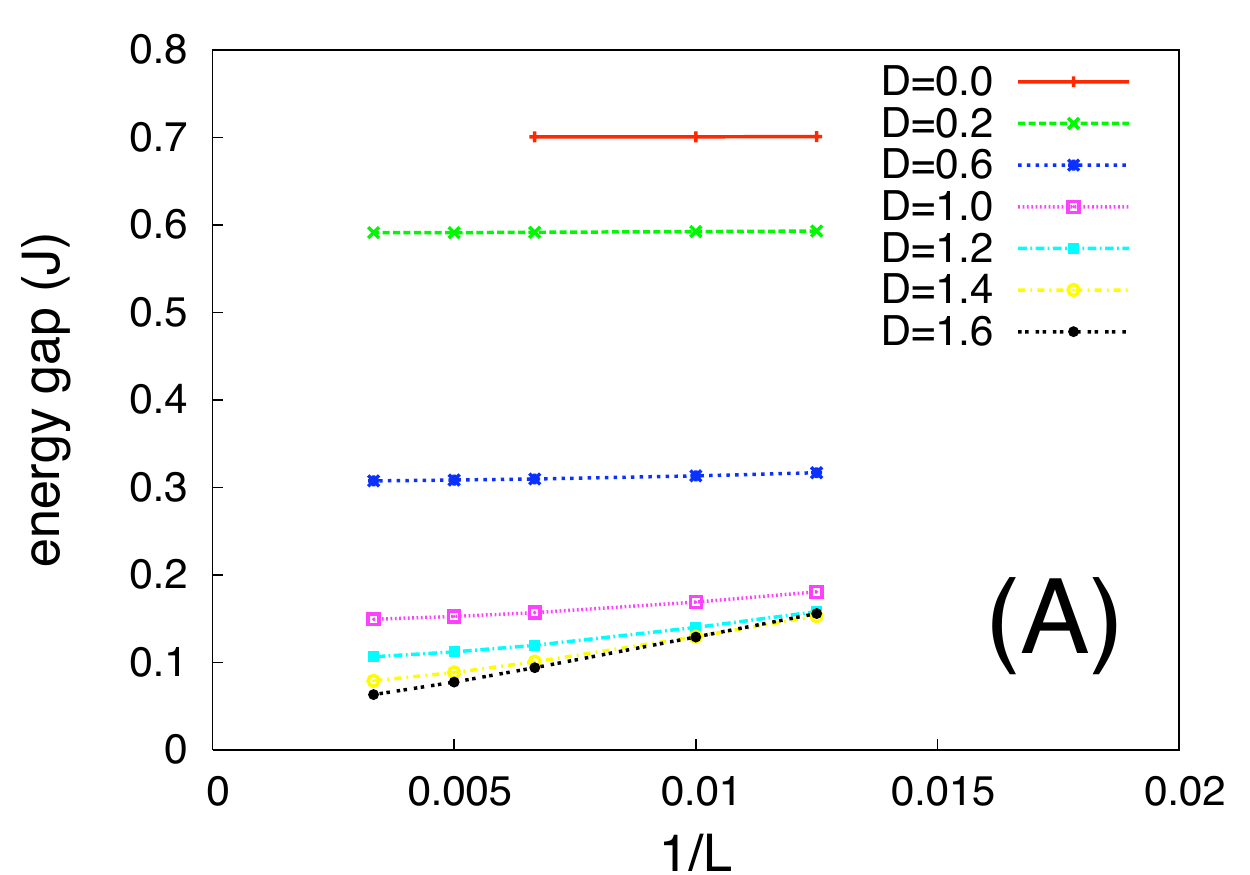}
\includegraphics[width=0.5\textwidth]{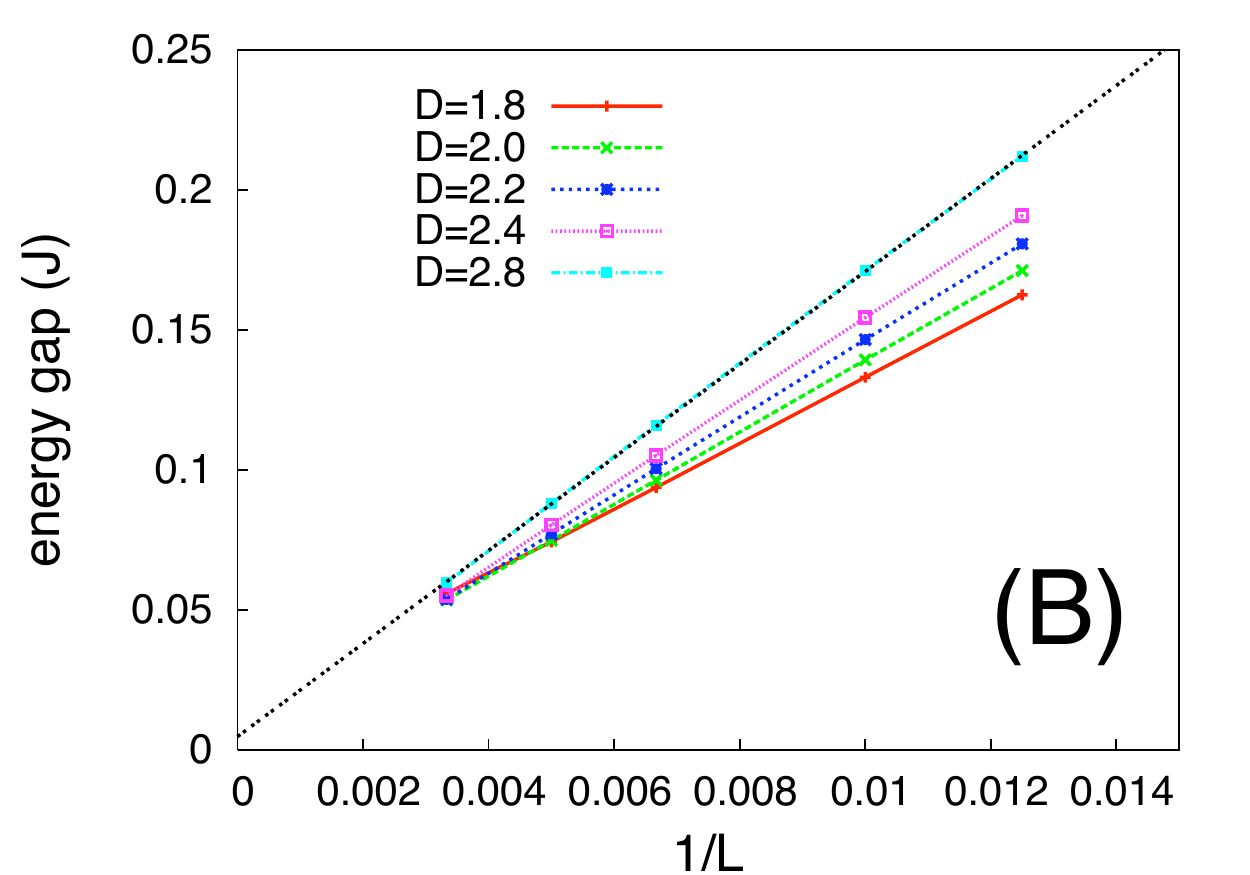}
\caption{(Color online)  The finite-sized scaling of the spin gap in the AKLT model.  The energy unit is $J$, which is taken to be one in the calculation.  (a) $D$ from 0 to 1.6.  (b) $D$ from 1.8 to 2.8.  The dash line for $D=2.8$ is the guide to eyes. }\label{Fig:aklt_scaling}
\end{figure}

\begin{figure}[htb]
\includegraphics[width=0.5\textwidth]{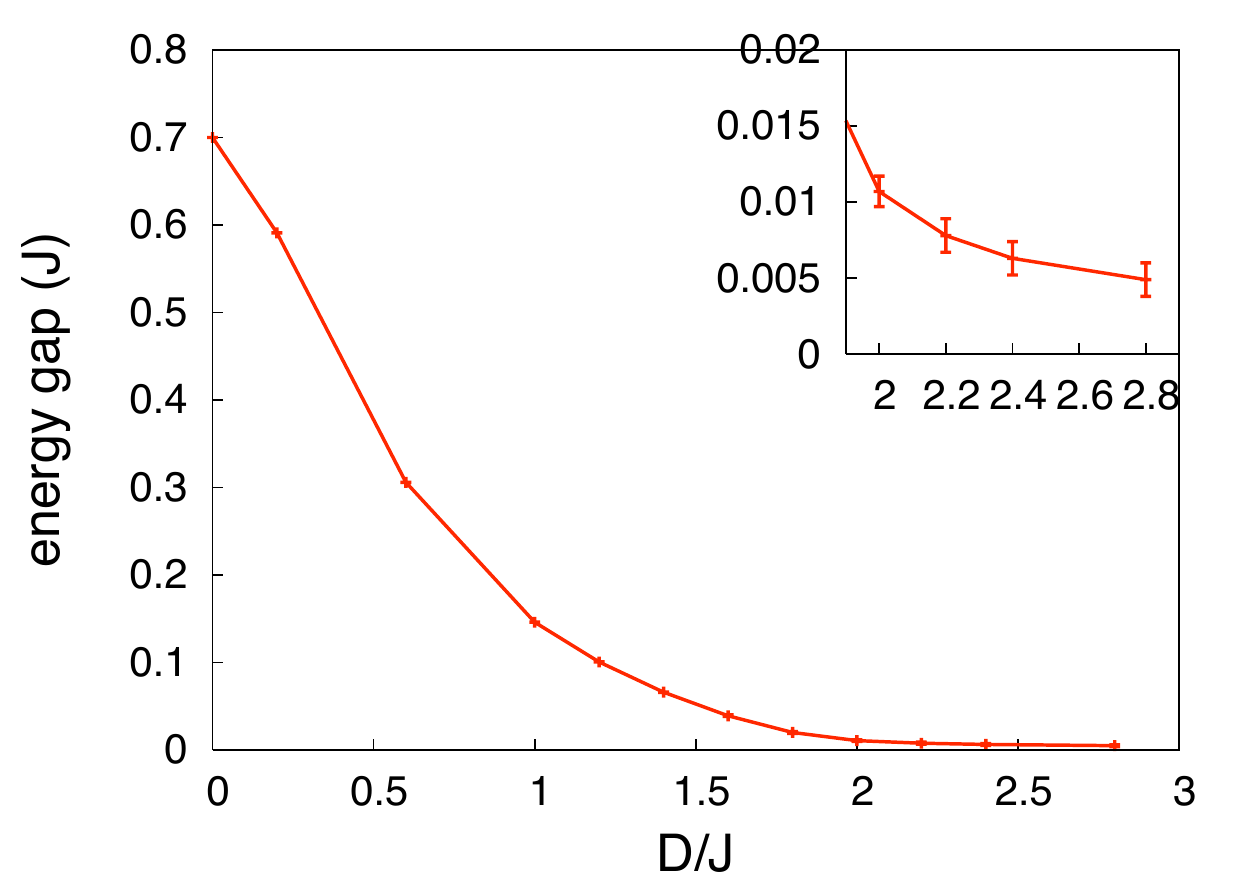}
\caption{(Color online) The spin gap as the function of $D/J$.  The gap function is a smooth function of $D/J$.  Although we only compute up to $D/J=2.8$, the tendency denies the quantum phase transition at finite $D/J$ as the Haldane gap case.}\label{Fig:aklt_gap}
\end{figure}

In Fig.(\ref{Fig:aklt_scaling}), we show the finite-sized scaling of the spin gap for different $D/J$.  Since the excitation of the VBS state is to break a singlet, we focus on the triplet excitations.  We extrapolate the scaling and obtain spin gap $\Delta_{\infty}$ in the thermodynamic limit for different $D/J$.  The result is shown in the Fig.(\ref{Fig:aklt_gap}).  As our DMRG accuracy is around $1\times 1.1\times10^{-3}$, we do not observe the spin gap to close up to $D/J\sim 3$.  It can be seen that the gap drops rapidly around $D/J\sim 0.6$ and is followed by a slow saturation, in which the gap remains finite and shows no evidence of the quantum phase transition at finite $D/J$.  Although it is not shown here, we have checked that it is indeed gapless for the pure DM model of spin one.  Together with the presented results, we conclude that the quantum phase transition does not occur until $D/J$ goes to infinity.

Our result demonstrates a nontrivial consistency between the Heisenberg model and the AKLT model.  That the VBS state in both systems are stable against the DM perturbation indicates that they are of the same kind.  Our results also imply the irrelevance to the Haldane gap of the $(\vec{S}_i\cdot\vec{S}_{i+1})^2$ term in the AKLT model, since the nearest-neighbored DM term can not be gauged away due to that term and the spin gap is still robust.  

\section{QD model}\label{section:QD}
In this section, we consider the QJ model given by Eq.(\ref{eq:QJ}) in the square lattice.\cite{Sandvik2007prl}  If $Q=0$, the ground state is the antiferromagnetic state.  If $J=0$, the ground state is shown to have the VBS order.  There is a quantum phase transition at $J^*/Q=0.04$ that both the Neel's order and the VBS order vanish, which is shown to be the second order phase transition with emergent global U(1) symmetry.  The small value of the $J^*/Q$ indicates that the VBS state is very vulnerable when competing with the antiferromagnetic state, that serves an good example to explain the current theoretical difficulty that it is very difficult to construct a pure local quantum spin model to host the VBS state as the ground state.  

Now, we would like to examine if this VBS state is also at the weak side when competing with the DM interaction?  In the transition metal oxides, it is quite common that the inversion symmetry is not exactly preserved.  Consequently, the DM interaction, regardless big or small, is not vanishing.  If the VBS state is also very vulnerable against the DM interaction, it will be a hard task for us to observe it experimentally.  To answer the question, let us take the QJ model plus the DM interaction and take $J=0$ as the example.  The Hamiltonian is given by the following
\begin{eqnarray}
\!\!H\! =\! -Q\!\!\!\!\sum_{<ijkl>}\!\!\!(\vec{S}_i\!\cdot\!\vec{S}_{j}\!-\!\frac{1}{4})(\vec{S}_k\!\cdot\!\vec{S}_{l}\!-\!\frac{1}{4})\!+\!\!\!\sum_{<ij>}\!\!\vec{D}_{ij}\!\cdot\! (\vec{S}_i\!\times\!\vec{S}_{j}). \label{eq:QD}
\end{eqnarray}
Let us call it the QD model and work in the square lattice.  To perform the calculation, the specific configuration of the $\vec{D}_{ij}$ is needed.  Here we borrow the $\vec{D}_{ij}$ from the case of the DM interaction in the perovskite La$_2$CuO$_4$\cite{Coffey1990prb}.  Although the ground state of La$_2$CuO$_4$ is not the VBS state, the configuration of the DM vector in this system is quite typical.  Readers should be clear that we are not studying the system of La$_2$CuO$_4$ here.

\begin{figure}[htb]
\includegraphics[width=0.4\textwidth]{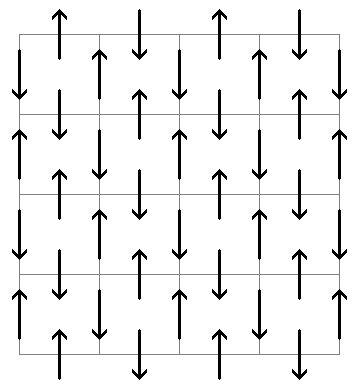}
\caption{(Color online)  The DM vector $\vec{D}_{ij}$ in the tetragonal phase in La$_2$CuO$_4$.  Here we only borrow the configuration of $\vec{D}_{ij}$ as a typical example to consider the the DM perturbation on the VBS state.}\label{Fig:DM_vector}
\end{figure}

In the La$_2$CuO$_4$, the DM interaction occurs in two different cases depending on the way of tilting of the CuO octahedra\cite{Coffey1990prb}.  Two situations happen in the orthorhombic and the tetragonal phases respectively.  Actually, they belong to the same kind of configurations, so we only study the tetragonal case where the DM vectors $\vec{D}_{ij}$ are given by Fig.(\ref{Fig:DM_vector}).  

We apply the Lanczos method to compute the spin gap of the Eq.(\ref{eq:QD}).  We consider the periodic boundary condition in the both $x$ and $y$ directions, and therefore the momentum is a good quantum number.  The sparse Hamiltonian is divided by the subspaces with the quantum numbers of the total magnetization $S$ and the total momentum $P$.  The ground state is always found in the $S=0$ and $P=0$ sector.  However, the first excited state is found either in the $S=0$ and $P=0$ sector or in the $S=1$ and $P=0$ sector in the calculation for different sizes of the system.  The first one is referred to the singlet excitation, and the latter one is the triplet excitation.  We expect that the first excitation is due to the broken bond of one valence bond and therefore should be in the triplet sector.  Indeed, we found that the first triplet excitation scales faster than the singlet one to have the lower energy.  Therefore, we look for the first excited state in the triplet sector.

\begin{figure}[htb]
\includegraphics[width=0.5\textwidth]{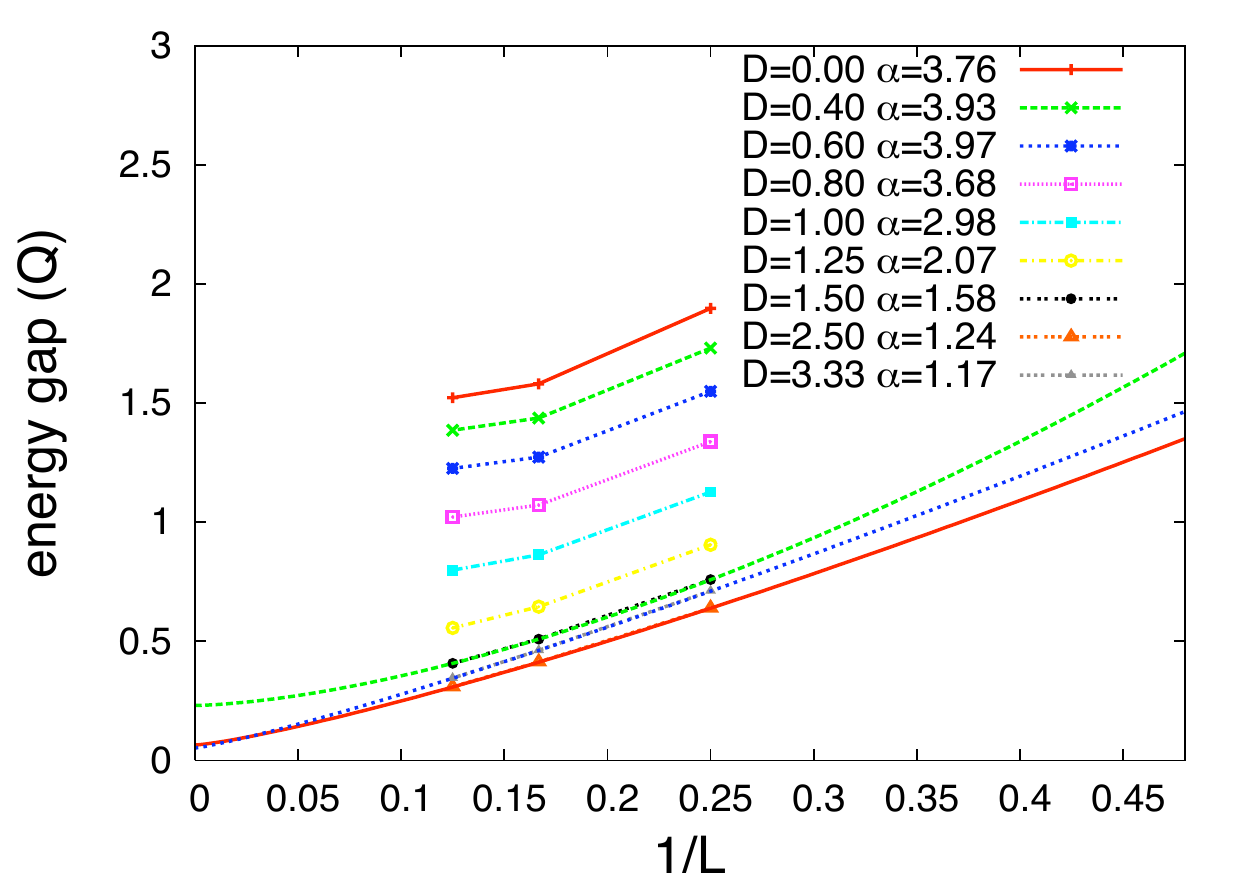}
\caption{(Color online)  The finite size scaling of $1/L$ of the spin gap as the function of $D/Q$.  The system size is $4\times L$.  We compute for $L = 4$, 6, and 8.  Both $x$ and the $y$ directions take the periodic boundary condition.  $\alpha$ is the fitting parameter so that the energy gap scales as $1/L^{\alpha}$}\label{Fig:qd_scaling}
\end{figure}

\begin{figure}[htb]
\includegraphics[width=0.5\textwidth]{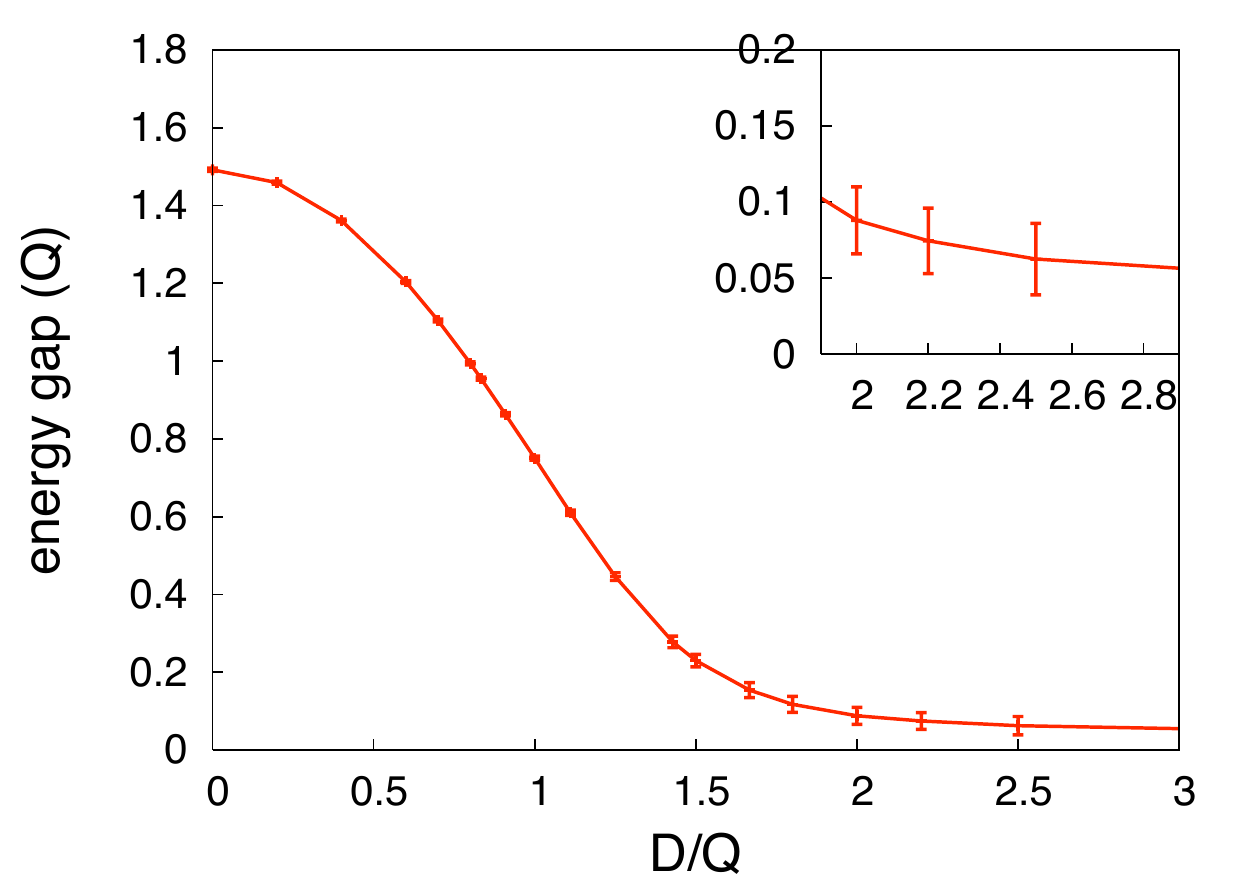}
\caption{(Color online)  The spin gap $\Delta_{\infty}$ as the function of $D/Q$.  The gap does not close at our largest computed D/Q = 3.33.}\label{Fig:qd_gap}
\end{figure}

Fig.(\ref{Fig:qd_scaling}) is the result of the finite-sized scaling as $1/L$ of the spin gap for different $D/Q$ values.  The system sizes are taken to be $4\times L$ with $L = 4$, 6, and 8.  We fit the scaling of the spin gap $\Delta(L)$ by
\begin{eqnarray}
\Delta(L) = \Delta_{\infty} + O(\frac{1}{L^{\alpha}}) \label{eq:qd_scaling}
\end{eqnarray}
where $\Delta_{\infty}$ is the value by taking $L$ to be infinity.  Although it is not shown here, we also compute the spin gap of the pure DM model, namely $Q=0$ in the QD model.  In this case, $\alpha =1$, and we found $\Delta_{\infty}(Q=0)=0$.  For finite $Q$, $\alpha > 1$ and change smoothly to $\alpha = 1$, indicating the lack of the quantum phase transition at finite $Q$.  We also compute the $\Delta_{\infty}$ as the function of $D/Q$ in the Fig.(\ref{Fig:qd_gap}).  While the error to compute the energy using the Lanczos method is given around the order of $10^{-6}Q$, the error bars in Fig.(\ref{Fig:qd_gap}) are mainly from fitting the finite-sized scaling.  

As the DM interaction is turned on, the spin gap reduces.  It ramps down rapidly around $D/Q\sim$ 1 but still remains finite after reaching the foot of the hill.  The result can be more accurate if we also do the finite-sized scaling along the $x$ direction.  Nonetheless, the result will change quantitatively but not qualitatively.  All values will become just smaller since the gap due to the finite size along the $x$ direction also reduces.  Here we compute up to $D/Q = 3.33$ and fail to observe the quantum phase transition.  From the tendency of the curve, we conclude that the gap does not close until $Q = 0$.

Let us briefly comment about the introduction of the $J-$term in the QD model, namely the QJD model.   Due to the reduction of the spin gap by the DM interaction, we can imagine that the quantum phase transition from the VBS state to the AFM state in the QJ model should happen at smaller $J^*$ in the QJD model.  However, whether or not the quantum phase transition is still of the second order type remains further study.  We remark that it is not an easy task, because the minus-sign problem returns in the QJD model, as the QJ model is free of that problem.  Unfortunately, it is not appropriate to apply the Lanczos method to study the QJD model.  Since it requires a larger system to see the finite-sized scaling to locate the transition point, the size that the Lanczos method can reach is not enough.
 
\section{Conclusion and Discussion}\label{section:CD}

In this paper, we have apply DMRG method and the Lanczos method to compute the spin gap in the AKLT model and the QD model with the DM interaction.  In both systems, the spin gap of the VBS state is stable against the DM perturbation.  The quantum phase transition does not occur until the pure DM model is reached.  Albeit the VBS state is difficult to obtain both experimentally and theoretically, our results show positively that the VBS state will not be destroyed by the DM interaction which is very common in the transition metal oxides or similar chemical families.  We understand that our result is different from the VBS state of the Heisenberg model in the kagome lattice, where at $D/J=0.1$ the transition from the VBS to the DM ground state occurs.  However, the ground state of the Heisenberg model in the kagome lattice has not reach consensus yet.  Moreover, the VBS state in the frustrated lattice could be very different from the one in the bipartite lattice.  Therefore, our results do not cause inconsistency with previous results.

To extend the current research, it is interesting to find out the nature of the quantum phase transition from the VBS state to the DM ground state.  As it is shown in the Ref[\cite{Neirotti1999prb}], it is the Kosterlitz-Thouless-type transition in the Heisenberg model in the spin chain.  Although we expect the same result in the AKLT model, the one in the QD model will be interesting.



This work is supported by NSC 97-2112-M-002-027-MY3 and the start-up grant from the National Taiwan University.


\end{document}